# Active mode locking of quantum cascade lasers operating in external ring cavity

D.G. Revin*[1], M. Hemingway[1], Y. Wang[2], J.W. Cockburn[1], A. Belyanin[2]

[1] *Physics and Astronomy Department, The University of Sheffield, S3 7RH, Sheffield, United Kingdom,*

[2] *Department of Physics and Astronomy, Texas A&M University, College Station, 77843 USA*

### Abstract

Stable ultrashort light pulses and frequency combs generated by mode-locked lasers have many important applications including high-resolution spectroscopy, fast chemical detection and identification, studies of ultrafast processes, and laser metrology. While compact mode-locked lasers emitting in the visible and near infrared range have revolutionized photonic technologies, the systems operating in the mid-infrared range where most gases have their strong absorption lines, are bulky and expensive and rely on nonlinear frequency down-conversion. Quantum cascade lasers are the most powerful and versatile compact light sources in the mid-infrared range, yet achieving their mode locked operation remains a challenge despite dedicated effort. Here we report the first demonstration of active mode locking of an external-cavity quantum cascade laser. The laser operates in the mode-locked regime at room temperature and over the full dynamic range of injection currents of a standard commercial laser chip.

1. Introduction

Quantum cascade lasers (QCLs) are based on electron intersubband transitions in coupled quantum wells and superlattices; they offer unique flexibility in engineering the emission wavelength and the width of the gain spectrum. Recently, free-running QCLs with octave-spanning lasing spectra [1] and a long-lived phase coherence of laser modes [2,3] were demonstrated and promising applications such as dual-comb spectroscopy were identified [4]. However, attempts to mode-lock mid-infrared QCLs have had limited success. Passive mode locking has been found to be impossible in QCLs. The reason lies in the fundamental difference between the dynamics of QCLs and virtually every other kind of laser. For the intersubband transitions, the ultrafast carrier relaxation time limits the gain recovery time to just a few picoseconds [5]. This is much shorter than the roundtrip time of 40-70 ps for a mid-infrared QCL with a typical cavity length of a few mm. Under these conditions the tails of the pulse experience unsaturated gain and therefore amplification, thus preventing the formation of stable mode-locked pulses.

Active mode locking (AML) by gain modulation at the period equal to the round-trip time was identified as the only viable route to QCL mode locking [6,7,8,9]. In this case a stable pulsed regime is achieved by periodically opening a window of net gain on a timescale that is short compared to the cavity round-trip time. In terahertz QCLs this could be achieved by launching a wave of microwave modulation of voltage [10]. In mid-infrared QCLs the only successful attempt of the AML was achieved in two-section monolithic cavity lasers with a so-called "super-diagonal" design with a significantly longer gain relaxation time, at the expense of overall device performance. The AML was observed only close to the threshold and only at cryogenic temperatures [6].

A more robust and potentially more practical approach to AML of QCLs is based on the use of QCLs operating in free-space external cavities [11]. In this case, one can utilize a standard high-performance QCL, modulate the whole laser chip, thus achieving deep gain modulation, while at the same time maintaining a very short gain window when compared to the round-trip time of the light along the external cavity. This extends the dynamic range of currents where the AML can be observed far above threshold. In addition, a long external cavity has round-trip frequencies reduced from tens of gigahertz in monolithic QCLs to hundreds of megahertz, which is much more convenient for gain or loss modulation and makes it easy to incorporate additional optical elements or a gas cell inside the cavity.

In this paper we demonstrate the first AML of an external-cavity QCL with a standard commercial laser chip. This is also the first demonstration of room-temperature mode locking in QCLs of any kind. The pulses maintain strict periodicity for at least $10^8$ roundtrips. Our results show that, contrary to widespread belief, neither short (of the order of 1 ps) gain recovery time nor the spatial hole burning in the gain medium prohibits AML of QCLs. The observed AML regime is very robust: it persists over the full dynamic range of the injection currents (from the laser threshold and up to the laser roll-over) and for all tested cavity lengths and at the modulation frequency range of ~80 – 400 MHz.

2. Experimental results

A 4.5 mm long, 9 µm wide λ ~ 5.25 µm buried heterostructure QCL Ref.[12] with both facets having anti-reflection (AR) coatings is used as a gain medium and driven with a sinusoidal current of variable frequency and amplitude. The optical and electrical arrangements of the free space external ring cavity set-up (Fig. 1a) are described in more details in Methods. When the laser chip is pumped with the current periodically modulated at the round-trip frequency of ~80.7 MHz for the ~3.8 m long cavity or its higher harmonics of up to ~403 MHz, the ring cavity QCL emits a periodic sequence of mode-locked pulses with the repetition rate equal to the frequency of the driving pulses (Fig. 1b). The shape and duration of these pulses are mainly defined by the ~2-ns rise time of the detector and to some extend by the bandwidth of the oscilloscope, and we believe that the actual duration of these pulses is likely to be much shorter as discussed below.

The emission pulses from the mode-locked QCL are well separated and have a constant amplitude and very high degree of periodicity. In order to assess the pulses phase stability, intermode beat (RF) spectra of the intensity are measured with an electronic spectrum analyzer. These spectra exhibit sharp, stable peaks at the modulation frequency and its harmonics. The full width at half maximum (FWHM) of less than 1.5 Hz has been estimated for the modulation with the round trip frequency or its second harmonic (Fig. 1c) and is likely to be limited by the 1 Hz resolution of the spectrum analyzer and the stability of the RF master generator. This means that the train of mode-locked emission pulses and the relative phases of the modes in the laser spectra remain stable for at least $10^8$ pulses. This is a much narrower beatnote than was ever reported for QCLs.

The average output power of the external ring cavity QCL driven with the current $I$ modulated at frequency near the round trip frequency and the amplitude from $I/I_{th}$ = 1.01 to $I/I_{th}$ = 1.7 is presented in Fig. 1d, where $I_{th}$ is the laser threshold current. Under resonant pumping the laser emission has a well-defined threshold current, which is the lowest when the repetition rate of the driving current pulses exactly matches the round trip frequency. If the modulation frequency is detuned from the resonance frequency, the mode-locked pulses are still generated but at higher threshold current values. It was found that the laser emits in both clockwise (CW) and counter clockwise (CCW) directions with identical characteristics but different intensities. The emission in the CCW direction is around 70% of the optical power for CW direction. We explain this asymmetry by slightly different reflectivities of the AR coatings at the laser chip facets. The experimental data presented here are for the CW direction only. The existence of the waves propagating in both CW and CCW directions means that the standing wave pattern can be formed inside the QCL chip, consequently leading to the formation of a transient population grating, or spatial hole burning (SHB). This effect has been shown to destroy mode locking in monolithic QCLs [13,14] since scattering on the population grating leads to proliferation of modes with uncorrelated phases. Our experimental results indicate that the active mode locking of external-cavity QCLs is possible even in the presence of SHB.

The maximum average output optical power out-coupled from the cavity using the beam-splitter increases linearly with the current and is almost 3 mW for the current amplitude of $I/I_{th}$ = 1.7. Beyond this point the emission intensity stops increasing, indicating that light-current roll-over conditions have been reached, in good agreement with the laser performance in the continuous wave or pulsed, non-resonant regimes. The frequency detuning range in which the laser operates is very narrow at threshold but increases rapidly for higher driving currents spreading around 3% both above and below the round-trip frequency. As expected, the output optical power reaches its maximum at the modulation frequency equal to the round-trip frequency for all values of current and rapidly decreases with detuning, when the window of net gain opens at times slightly out of phase with the round-trip time of the laser light.

The emission spectra of the laser change dramatically with modulation frequency detuning from the resonance. At the round-trip frequency and slightly below it (the data marked in green color in Fig. 1d) the emission spectra are very narrow, resolution limited, single peak, with FWHM of ~0.45 cm$^{-1}$, containing about 170 longitudinal modes of the ring cavity resonator. At the frequencies slightly above the resonance the spectra have several peaks with energy positions and intensities which abruptly jump/shift with slight variation in the pumping frequency (the data marked in red color). With further detuning from resonance, both above and below (the data marked in blue color), the emission spectra abruptly become very broad, with FWHM of up to 30 cm$^{-1}$, containing more than 12000 modes of the ring cavity. These broad spectra have relatively uniform shapes with only dips corresponding to the water vapor absorption lines. Typical single peak, multi-peak and broad spectra are presented in Fig. 1e. More detailed dependence of the spectra on the modulation frequency is illustrated in Fig. 1f for the current value of $I/I_{th}$ = 1.15. The frequency range areas marked as I, II and III correspond to the narrow single peak, multi-peak emission and the broad spectra, respectively. The origin of the narrow spectra in the vicinity of the resonance and the broadening of the spectra with modulation frequency detuning are discussed in more details in the theoretical modelling part.

Very similar dependence on the frequency detuning and, more importantly, the existence of the same types of the emission spectra are present if the driving current is modulated at the second and the third harmonics of the round-trip frequency (~ 161 and 242 MHz). Emission has also been detected for the modulation at the fourth and the fifth harmonics (~ 323 and 403 MHz). However, the operation of the interface driving board becomes much less efficient at higher frequencies. Active mode locking at higher harmonics yields shorter pulses at higher repetition rates for any given cavity length.

The maximum out-coupled peak optical power from the actively mode-locked QCL is estimated to be ~12 mW if the emission pulses are indeed 3-4 ns long. However, it is likely that this duration is limited by the resolution of the detector, since the same pulses are observed for all types of the emission spectra, both narrow and broad ones. The width of the observed narrow peak spectra under modulation near the round-trip frequency is 0.45 $cm^{-1}$ which should correspond to pulses of 75 ps duration and with a peak power of several hundred mW, assuming perfect mode locking. To confirm the pulse duration one would have to use an ultrafast QWIP or interferometry autocorrelation technique. Such measurements will be the subject of future research.

In order to control the central emission wavelength of the mode-locked ring cavity QCL for potential applications in spectroscopy, one of the external flat mirrors in the ring cavity (Fig. 1a) was replaced by a 300 lines/mm diffraction grating. The resulting length of the aligned cavity turned out to be a bit shorter: around 3.4 m, which corresponds to the round trip frequency of ~88.2 MHz. The use of the grating strongly limits the emission wavelength range of the mode locked QCL. The QCL is forced to operate only near the wavelength defined by the incident angle of the diffraction grating for all pumping frequencies near resonance (Fig. 2a). Very narrow spectra of FWHM ~ 0.45 cm$^{-1}$ are observed for all detuning frequencies. Similar to the case of an external cavity with no grating, there are three distinctively different frequency ranges of laser operation. However, the differences are now not in the emission spectra (limited now by the diffraction grating) but in the intermode beat spectra of the AML pulses (Fig. 2b). At the resonance frequency, and slightly below it, the RF spectra have very narrow and stable peaks (Fig. 2a, range I) at the modulation frequency and its higher harmonics. Right above the resonant frequency the RF spectra consist of several additional peaks near the pumping frequency and its harmonics corresponding to an amplitude modulation and resulting in a very small emission wavelength jittering (Fig. 2a, range II). For larger detuning frequencies (range III) the RF spectra start to develop a pedestal which gradually broadens with increasing detuning, indicating the reduced pulse amplitude stability and phase coherence. The rotation of the diffraction grating leads to the emission wavelength tuning in the range of ~150 cm$^{-1}$, very close to the FWHM of ~160 cm$^{-1}$ of the electroluminescence spectrum, measured for this AR coated QCL under non-resonant pulsed pumping without feedback from external mirrors (Fig. 2c).

3. **Theoretical modeling and discussion**

We model AML in an external cavity QCL with space- and time-domain simulations using a four-subband model for the QCL active region (see Methods). Figures 3 a and b show examples of the absolute values of the CW and CCW-propagating laser fields and their spectra when the bias is sinusoidally modulated with the frequency exactly at the resonance with the round-trip frequency. The time and the frequency in all theoretical plots are normalized by the round-trip time $T_{round}$ in an external cavity of 1.29 m. The DC level of the bias is close to the threshold and the maximum amplitude of the current is equal to $1.77 \times I_{th}$. In this example the high intensity pulses are present in both directions, which should lead to potentially strong SHB effects. However, the comparison of the CW and CCW pulses in Fig. 3a reveals an interesting feature: the pulses propagating in the opposite directions experience a self-consistent shift in their shape and relative phase of propagation which allows them to avoid an overlap in the gain medium. This mutual avoidance maximizes the gain available to each pulse and is entirely the result of their interaction through the coherent population grating, i.e. the SHB effect. The tendency of laser emission to avoid an overlap in time, space, or spectral domain in order to "feed" from unsaturated gain regions is of course a generic property of any laser. Since QCLs have a homogeneously broadened gain transition, the only mechanism which could facilitate interaction between the pulses is SHB. If the SHB is "turned off" by forcing the population grating amplitude to be equal to zero at all times, the pulses will have identical shape and completely overlap in the gain medium.

The pulse sides appear to be very sharp because the rise and fall times of the pulses are controlled by the gain recovery timescale which is three orders of magnitude shorter than the round-trip time.

Since the pulses do not overlap in the gain medium, there is no mode proliferation with random phases due to SHB and the pulses are essentially Fourier-transform limited, with narrow spectra shown in Fig. 3b. The pulse shapes may vary from run to run due to the presence of fluctuations and spontaneous emission noise in our model, but this avoidance feature is present whenever lasing occurs in both propagating directions. In most runs lasing in one direction eventually dies out over a long time scale of many hundreds or thousands of round-trips. In the experiment the two-directional lasing regime with narrow spectra is more stable.

The above physical picture of the pulse mutual avoidance and the resulting narrow spectra exists exclusively in the close vicinity of the resonance between the modulation period and the round-trip time. When the detuning is introduced, the pulses are not able to avoid each other in the active medium. Their mutual scattering on the coherent population grating leads to proliferation of longitudinal modes. The spectra experience a dramatic broadening to many thousands of external cavity modes and the pulses develop a substructure in time domain, with ultrafast oscillations on the picosecond timescale. This regime is illustrated in Figs. 3c,d for the case of modulation frequency 1% lower compared to the round-trip frequency and the modulation amplitude corresponding to the maximum current amplitude of $1.15 \times I_{th}$. With further detuning in any direction the qualitative pattern remains the same but the pulse amplitude decreases until the laser ceases to start.

The theoretical findings of the pulses with the narrow spectra in the vicinity of the resonance followed by the dramatic expansion of the spectra and eventual shutdown of the laser with modulation frequency detuning are in good qualitative agreement with the experimental data. The modeling also shows that one can achieve much shorter pulses and broader phase-locked frequency combs by modulating the pumping with shorter and sharper pulses instead of the sinusoidal modulation. Fig. 3e presents the laser output electric field and the corresponding spectrum when the applied bias is a sequence of Gaussian pulses with zero DC offset and $0.2 \times T_{round}$ duration at the $1/e$ level. The pulse repetition rate is exactly equal to the round-trip frequency and the peak current amplitude is $1.77 \times I_{th}$. In this case of narrower pumping pulses the uni-directional regime is achieved much faster. The resulting pulses, which propagate only in one direction, are not affected by the SHB effects; they have a nice symmetric shape and Fourier-transform limited spectra. The surviving lasing direction fluctuates from CW to CCW randomly from run to run. For practical purposes one can introduce an optical isolator into the cavity to select one definite lasing direction.

When the modulation period is slightly detuned from resonance the lasing fields in both directions can survive. Similar to the case of sinusoidal modulation, pulses interact strongly in the laser chip due to scattering on the coherent population grating, resulting in the ultrafast substructure and much broader spectra. This regime shows long-term instabilities: the pulses demonstrate amplitude modulation which fluctuates from run to run, and in many cases the lasing dies out in one direction. Again, the use of an optical isolator would be beneficial.

With decreasing duration of the driving pulses the output laser pulses become shorter. The simulations for the varying width of the voltage pulses have revealed that the shortest output pulse duration is limited, in order of magnitude, by the time it takes the pulse to travel along the laser chip, which is of the order of 10-45 ps for a 1 - 4.5 mm chip. Much shorter pulses will experience gain only in a small part of the laser chip. The corresponding voltage pulses will have the duration of about 30-60 ps which is at the limit of capabilities of modern electronics.

Since SHB does not destroy mode locking when counter-propagating pulses overlap in the active medium, one should expect that the AML is also possible in a Fabry-Perot cavity configuration. Indeed the simulations done for a Fabry-Perot cavity yield the results qualitatively similar to those for a ring cavity. They will be reported elsewhere, together with experimental data.

**Conclusions**

Robust, stable generation of actively mode locked pulses has been demonstrated from a room temperature λ ~ 5 μm external-cavity quantum cascade laser driven by the periodically modulated current with the modulation frequency close to the round-trip cavity frequency of 80.7 MHz or its harmonics. The active mode locking has been achieved at driving current amplitudes covering the entire dynamic range of the laser. An ultra-narrow RF beatnote linewidth below 1.5 Hz was recorded. Introduction of a diffraction grating into the ring cavity provides a single peak emission tuning range of ~150 cm$^{-1}$. The theoretical simulations are in good agreement with the observed dynamics and shed light on the physical mechanism of active mode locking in QCLs.

**Methods**

The free space X-shaped external ring cavity QCL set-up (Fig. 1a) is very similar to one described in Ref.[15] and includes "Alpes Lasers" QCL, two AR coated 4 mm focal length chalcogenide lenses, four flat uncoated silver mirrors and a 0.5 mm thick CaF$_2$ beam-splitter AR coated on one side to suppress FP interference in it and to provide ~5% reflectivity to out-couple the light. The temperature of the laser chip submount is kept at 15 $^0$C. The length of the cavity of 3.4-3.8 m has been chosen in order to clearly resolve the periodic emission pulses expected at the fundamental round trip time (~11-12 ns) and at its higher harmonics, by a stand-alone "Vigo" mercury cadmium telluride (MCT) detector which has a rise time of ~2 ns. An FTIR "Bruker IFS66" spectrometer was utilized for spectral recordings and a thermopile detector was used for average optical power measurements. "Avtech" pulse generator providing 1 µs - 1 ms long pulses with up to a 50 % duty cycle was used to align the ring cavity set-up prior the resonant pumping experiments. The laser threshold current in the pulsed non-resonant regime is ~400 mA with the roll-over current at around 750 mA. The remaining reflectivity of less than 2% of the AR coated laser chip facets has reduced the FP feedback from these facets to such an extent that the laser no longer operates as a free running pulsed laser device outside the external cavity. The ring cavity QCL was also tested in the continuous wave regime demonstrating the performance characteristics similar to those reported in Ref.[15]. The sinusoidal electrical signal with variable frequency and amplitude is supplied by an "Agilent" RF master generator. The output from the generator is connected to the specially designed interface board acting as a broad band (~60 - 400 MHz) power amplifier with adjustable amplification based on a high frequency high power LDMOS FET transistor. The interface board is designed to provide negative current pulses of up to ~1 A on 50 Ohms impedance load, to have zero DC bias offset and strong rejection of the signal with the positive polarity. Because of the interface board design and the voltage dependent resistance of the QCL the resulting shape of the current modulation swinging from zero value to the operating values is found to be not sinusoidal. On the other hand, having the driving current pulses with sharper and narrower peaks may be considered advantageous for this type of experiments. To monitor the shape and the amplitude of the driving current pulses a ~3 ns rise time inductive current probe has been introduced directly before the QCL chip. The outputs from the current probe and the fast MCT detector are connected to an oscilloscope with 300 MHz bandwidth. Due to the high impedance mismatch with the QCL which results in the strong reflected RF signal, the shape of the driving current pulses observed on the oscilloscope does not necessarily represent the current pulses applied directly to the QCL chip. For that reason the exact measurements of the current values are unreliable and all current amplitudes are therefore given only as a $I/I_{th}$ ratio.

The theoretical approach for the theoretical model of AML in an external cavity QCL includes full spatiotemporal dynamics of the laser field, intersubband polarization, and electron populations within coupled Maxwell and density-matrix equations. The active region model is described in detail in Ref.[9]. We use a comprehensive model of the QCL active region and transport which includes resonant tunneling injection, electron distribution over in-plane k-vectors, and space charge. We also perform time-domain and space-domain simulations of the propagating field, in which we can directly follow the pulse formation in time. Note that previous studies such as [13] and [14] used the frequency-domain modal approach and a two- or three-level laser model without current injection. We modified the field propagation model as compared to [9] to include appropriate geometry and boundary conditions for an external cavity. Ideal AR coatings on both facets of the laser and 5% out-coupling from the external cavity for both directions are assumed for all presented simulations. It was found that additional residual facet reflectivities of 1% have negligible effect on the results. A shorter external cavity of 1.29 m was used to reduce calculation times. The results normalized by the cavity length and the round-trip time are not sensitive to the increase in the cavity length to 3.8 m. The input parameter in the model is the applied bias voltage and not the current, which is calculated self-consistently taking into account electron distributions in the subbands. Therefore the current response to a sinusoidal modulation of the bias is non-sinusoidal, but the nonlinearity in the response is small for the used modulation amplitudes.

**References**

1. Rosch, M., Scalari, G., Beck, M. & Faist, J. Octave-spanning semiconductor laser. *Nature Photon.*, **9**, 42-47 (2015).

2. Hugi, A., Villares, G., Blaser, S., Liu, H.C., & Faist J. Mid-infrared frequency comb based on a quantum cascade laser. *Nature.* **492**, 229-233 (2012).

3. Burghoff, D. *et al.* Terahertz laser frequency combs. *Nature Photon.* 8, 462 (2014).

4. Villares, G., Hugi, A., Blaser, S., & Faist, J. Dual-comb spectroscopy based on quantum-cascade-laser frequency combs, *Nature Comm.* **5**, 5192 (2014).

5. Choi, H*. et al.* Femtosecond dynamics of resonant tunneling and superlattice relaxation in quantum cascade lasers. *Appl. Phys. Lett.* **92**(12), 122114–122117 (2008).

6. Wang, C.Y. *et al.* Mode-locked pulses from mid-infrared quantum cascade lasers. *Opt. Exp.* **17**, 12929-12943 (2009).

7. Barbieri, S. *et al.* Coherent sampling of active mode-locked terahertz quantum cascade lasers and frequency synthesis. *Nature Photon.* **5**, 306 (2011).

8. Wojcik, A.K. *et al.* Generation of picosecond pulses and frequency combs in actively mode locked external ring cavity quantum cascade lasers. *Appl. Phys. Lett.* **103**, 231102 (2013).

9. Wang, Y. & Belyanin, A. Active mode-locking of mid-infrared quantum cascade lasers with short gain recovery time. *Opt. Exp.* **23**(4), 4173-4185 (2015).

10. Dhillon, S. Terahertz pulse generation from quantum cascade lasers, talk at IWFCTA 2015, Vienna, Austria; 11.09.2015.

11. Malara, P. *et al.* External ring-cavity quantum cascade lasers. *Appl. Phys. Lett.* **102**, 141105 (2013).

12. Blaser, S. *et al.* Low-consumption (below 2 W) continuous-wave single mode quantum-cascade lasers grown by metal-organic vapour phase epitaxy. *Electron. Lett.* **43**, N 22, 1201–1202, (2007).

13. Gordon, A. *et al.* Multimode regimes in quantum cascade lasers: from coherent instabilities to spatial hole burning. *Phys. Rev. A*, **77**, 053804 (2008).

14. Gkortsas, V.M. *et al.* Dynamics of actively mode-locked quantum cascade lasers. *Opt. Exp.* **18**(13), 13616-13630, (2010).

15. Revin, D.G. *et al.* Continuous wave room temperature external ring cavity quantum cascade laser. *Appl. Phys. Lett.* **106**, 261102 (2015).

**Acknowledgments**

D.G.R, M.H. and J.W.C. acknowledge support from the Seventh Framework Programme of the European Union, FP7-PEOPLE-2011-IAPP "QUANTATEC" Grant No. 286409. Y.W. and A.B. acknowledge the support by the Air Force Office for Scientific Research through grant FA9550-15-1-0153 and by NSF Grant EEC-0540832 (MIRTHE ERC). M.H. also acknowledges support from the EPSRC studentship.

**Contributions**

D.G.R. designed and carried out the experiments, with assistance from M.H. M.H. designed and fabricated the RF interface QCL driving board. Y.W. and A.B. performed the numerical simulations. D.G.R. and A.B. analysed the data and wrote the manuscript. The project was organized and coordinated by J.W.C. and D.G.R.

**Competing financial interests**

The authors declare no competing financial interests.

**Corresponding authors**

Correspondence to: D.G. Revin or A. Belyanin

**Figure captions**

**Fig.1** Performance of the actively mode locked free-space external cavity QCL. **a,** Optical setup. QCL, quantum cascade laser; L, aspheric lenses; M, mirrors; BS, beamsplitter; MCT, detector; DG, diffraction grating; CW, clockwise direction; CCW, counter clockwise direction. **b**, Periodic emission pulses for the modulation at the round-trip frequency of 80.7 MHz. **c**, Intermode beat spectra for the emission pulses with the current modulated near round-trip frequency and its second harmonic of 161.4 MHz. **d**, Average output power of the external ring cavity QCL driven near the resonant frequency with the current from $I/I_{th}$ =1.01 to $I/I_{th}$ = 1.7 and the incremental step of $0.1 \times I_{th}$ from curve to curve. **e**, Typical emission spectra for various ranges of the pumping frequencies. **f**, Emission spectra of the external ring cavity QCL driven

with the current value of $I/I_{th}$ = 1.15 and modulated near the resonant frequency. Transmission spectrum for water in the air is shown as a reference.

**Fig.2** Emission spectra of the actively mode locked external ring cavity QCL with a diffraction grating. **a,** The laser is modulated near the resonant frequency and driven with $I/I_{th}$ = 1.15. The angle of the diffraction grating is fixed. The frequency ranges I, II and III correspond to different shapes of the intermode beat spectra. **b**, Typical intermode beat spectra of the emission pulses for various ranges of the modulation frequencies. **c,** Set of the narrow single peak spectra with FWHM ~ 0.45 cm$^{-1}$ at $I/I_{th}$ = 1.6 demonstrating the wavelength tuning by rotating the diffraction grating (in range I for the modulation frequencies). For comparison: electroluminescence (EL), measured with 20 cm$^{-1}$ resolution.

**Fig.3**. Calculated laser output under resonant modulation of the bias of a QCL in a ring cavity configuration. The time and the frequency are normalized by the round-trip time $T_{round}$ in an external cavity. **a**, Absolute values of the CW and CCW propagating electric fields $|E|$ for a sinusoidal modulation of the bias with the modulation period equal to $T_{round}$. **b**, Spectra of the electric fields shown in Fig. 3 a. **c**, Absolute values of the electric fields for a sinusoidal modulation of bias with the modulation period equal to 1.01×$T_{round}$. **d**, Spectra of the electric fields shown in Fig. 1 c . **e,** Absolute value of the electric field and its spectrum when the applied bias is a sequence of Gaussian pulses with the modulation period equal to $T_{round}$ and duration of 0.2×$T_{round}$.

Fig.1

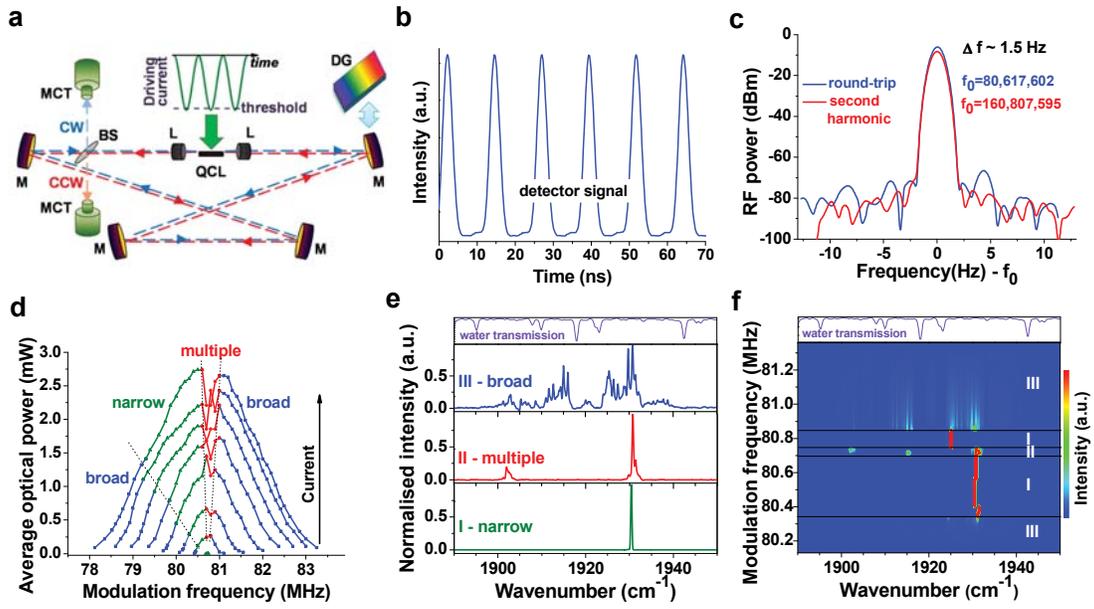

Fig.2

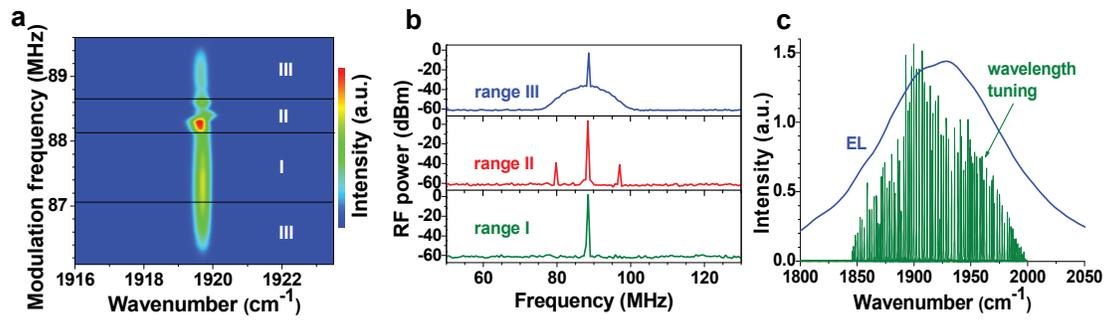

Fig.3

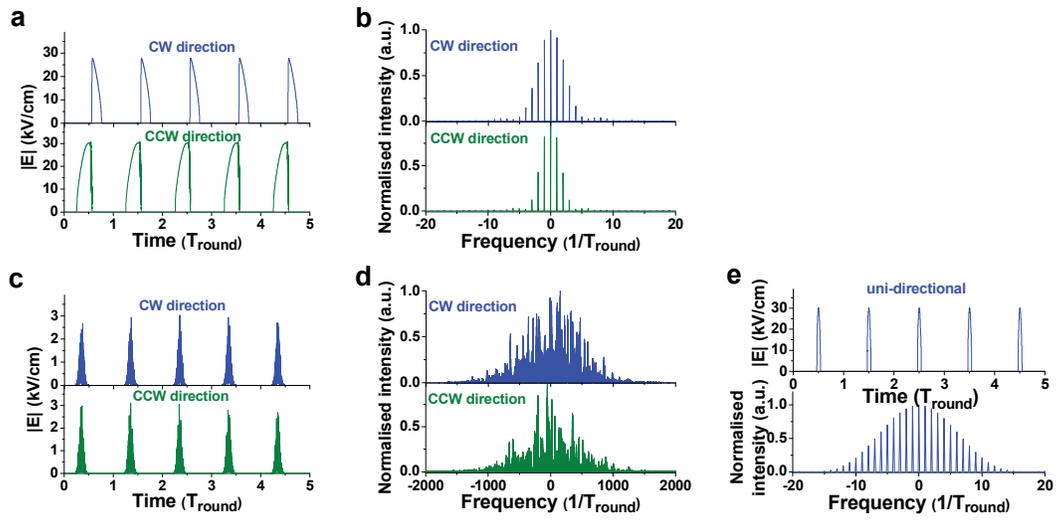